\documentclass{article}

\usepackage{amssymb}

\usepackage{epsfig}

\usepackage{lscape}

\usepackage{graphicx,psfrag,amsmath}

\begin{document}

\def\beq#1\eeq{\begin{equation}#1\end{equation}}
\def\beql#1#2\eeql{\begin{equation}\label{#1}#2\end{equation}}

\def\bea#1\eea{\begin{eqnarray}#1\end{eqnarray}}
\def\beal#1#2\eeal{\begin{eqnarray}\label{#1}#2\end{eqnarray}}

\newcommand{\Z}{{\mathbb Z}}
\newcommand{\N}{{\mathbb N}}
\newcommand{\C}{{\mathbb C}}
\newcommand{\Cs}{{\mathbb C}^{*}}
\newcommand{\R}{{\mathbb R}}
\newcommand{\intT}{\int_{[-\pi,\pi]^2}dt_1dt_2}
\newcommand{\cC}{{\mathcal C}}
\newcommand{\cI}{{\mathcal I}}
\newcommand{\cN}{{\mathcal N}}
\newcommand{\cE}{{\mathcal E}}
\newcommand{\cA}{{\mathcal A}}
\newcommand{\xdT}{\dot{{\bf x}}^T}
\newcommand{\bDe}{{\bf \Delta}}

\def\ket#1{\left| #1\right\rangle }
\def\bra#1{\left\langle #1\right| }
\def\braket#1#2{\left\langle #1\vphantom{#2}
  \right. \kern-2.5pt\left| #2\vphantom{#1}\right\rangle }
\newcommand{\gme}[3]{\bra{#1}#3\ket{#2}}
\newcommand{\ome}[2]{\gme{#1}{#2}{\mathcal{O}}}
\newcommand{\spr}[2]{\braket{#1}{#2}}
\newcommand{\eq}[1]{Eq.\,(\ref{#1})}
\newcommand{\xp}[1]{e^{#1}}

\def\limfunc#1{\mathop{\rm #1}}
\def\Tr{\limfunc{Tr}}

\def\dr{detector }
\def\drn{detector}
\def\dtn{detection }
\def\dtnn{detection}

\def\pho{photon }
\def\phon{photon}
\def\phos{photons }
\def\phosn{photons}
\def\mmt{measurement }
\def\an{amplitude}
\def\a{amplitude }
\def\co{coherence }
\def\con{coherence}

\def\st{state }
\def\stn{state}
\def\sts{states }
\def\stsn{states}

\def\cow{"collapse of the wavefunction"}
\def\de{decoherence }
\def\den{decoherence}
\def\dm{density matrix }
\def\dmn{density matrix}

\newcommand{\mop}{\cal O }
\newcommand{\dt}{{d\over dt}}
\def\qm{quantum mechanics }
\def\qms{quantum mechanics }
\def\qml{quantum mechanical }

\def\qmn{quantum mechanics}
\def\mmtn{measurement}
\def\pow{preparation of the wavefunction }

\def\me{ L.~Stodolsky }
\def\T{temperature }
\def\Tn{temperature}
\def\t{time }
\def\tn{time}
\def\wfs{wavefunctions }
\def\wf{wavefunction }
\def\wfn{wavefunction} 
\def\wfsn{wavefunctions}
\def\wvp{wavepacket }
\def\pa{probability amplitude } 
\def\sy{system } 
\def\sys{systems }
\def\syn{system} 
\def\sysn{systems} 
\def\ha{hamiltonian }
\def\han{hamiltonian}
\def\rh{$\rho$ }
\def\rhn{$\rho$}
\def\op{$\cal O$ }
\def\opn{$\cal O$}
\def\yy{energy }
\def\yyn{energy}
\def\yys{energies }
\def\yysn{energies}
\def\pz{$\bf P$ }
\def\pzn{$\bf P$}
\def\pl{particle }
\def\pls{particles }
\def\pln{particle}
\def\plsn{particles}

\def\plz{polarization  }
\def\plzs{polarizations }
\def\plzn{polarization}
\def\plzsn{polarizations}

\def\sctg{scattering }
\def\sctgn{scattering}

\def\prob{probability }
\def\probn{probability}

\def\om{\omega} 

\def\hf{\tfrac{1}{2}}

\def\zz{neutrino }
\def\zzn{neutrino}
\def\zzs{neutrinos }
\def\zzsn{neutrinos}

\def\zn{neutron }
\def\znn{neutron}
\def\zns{neutrons }
\def\znsn{neutrons}

\def\csss{cross section }
\def\csssn{cross section}

\def\vhe{very high energy }
\def\vhen{very high energy}

\def\mult{multiplicity }
\def\multn{multiplicity}

\def\bd{`black disc' }

\def\ed{`edge' }
\def\ed{`edge'}

\title{Evidence for a Energy-Invariant `Edge' in Proton-Proton
Scattering
at Very High Energies}

\author{ Martin M. Block\\
 Department of Physics and Astronomy \\
   Northwestern University,\    Evanston, Illinois 60208, USA \\
\\
Loyal Durand\footnote{Present address: 415 Pearl Court, Aspen, CO
81611}\\
Department of Physics, University of Wisconsin-Madison\\ 
Madison, Wisconsin  53706, USA
\\
\\
Francis Halzen\\
Wisconsin IceCube Particle Astrophysics Center and Department
of Physics,\\
University of Wisconsin-Madison\\
 Madison,  Wisconsin 53706, USA
\\
\\
Leo Stodolsky\footnote{Corresponding author: les@mpp.mpg.de} \\
Max-Planck-Institut f\"ur Physik
(Werner-Heisenberg-Institut)\\
F\"ohringer Ring 6, 80805 M\"unchen, Germany\\
\\
Thomas  J. Weiler\\
Department of Physics and Astronomy, Vanderbilt University,\\
Nashville, Tennesee
37235 USA
\\
\\
Aspen Working Group on Very High Energy Interactions, August 2014}

\maketitle

\begin{abstract}
Accurate fits to $pp$  and $\bar pp$  \csss data up to Tevatron
energies, incorporating the constraints imposed by 
analyticity and unitarity, successfully predict
 the results of recent LHC and cosmic ray measurements, and suggest
that the \csssn s approach  a black disc limit 
asymptotically. The approach to the limit is, however, very slow.
We present  a simple  geometric picture 
which explains  these features in a natural way. A \bd of
logarithmically growing radius is
supplemented by a soft `edge' whose properties are invariant with
\yyn. The constancy of the edge  results in the prediction that the
quantity $(\sigma^{TOT}-2\sigma^{El})/\surd\sigma^{TOT}$
approaches a constant at high \yyn. Using the existing fits, this
prediction  appears to 
be verified.  The value of the limiting constant allows an estimate
of
the thickness of the edge, which turns out to be on the order of
$1\,{\rm fm}$. One thus arrives at a picture where the
proton-proton scattering at lower
\yys is dominated by what becomes the  edge, while at higher \yys
it is dominated by the disc. The crossover between the two
regimes is only at  
$\surd s\geq $ 10 TeV, accounting for the slow approach to 
asymptotic behavior. Some questions as to the nature of the edge
are
discussed. 
\end{abstract}

\section{Introduction}

 Since the earliest days of high \yy physics,  questions 
and speculations have been raised about the ultimate nature and
geometric form of
very high \yy
elementary \pl interactions. A Yukawa-like  
mass-energy density
around the proton plus the assumption of a strongly
energy-dependent
interaction strength
leads to an effective radius of interaction $R\sim \ln s$ and thus
a \csss
$\sim R^2\sim
\ln^2s$ \cite{hbg}. A model with diffusion in the transverse
dimensions leads
to a similar conclusion, with additionally a relation between the
\csss and
the \mult which appears to be satisfied \cite{bl}. Analyticity
arguments lead  to the conclusion that a  $\ln^2 s$ growth of the
cross section is in fact the 
most rapid allowable   behavior \cite{frois,Martin}.

 Indeed, the very high \yy $pp$ and $\bar{p}p$ cross sections
approach a $\sigma\sim \ln^2s$ form at the
highest \yys as shown 
 by the fits in \cite{bh0}. These included data up to Tevatron
energies of $\sim 1.8$ TeV, and {\em accurately predicted}
both  the LHC and
cosmic ray experimental results \cite{bh1,bh}.
It is found that both the total \csss and the elastic \csss
have a leading $\ln^2s$ behavior. Furthermore the ratio of the
coefficients of
these terms are closely 2:1, as would be expected in  a black-disc
picture.   
However, in these fits the $\ln^2s$ terms are not totally dominant,
even at LHC
\yysn, indicating a very slow approach to ``asymptopia''.

   Here we  present a picture of the scattering in which these
features arise
in a natural way, and a test which seems to confirm the model and
to determine some of its parameters.

\section{Disc Plus Edge Model}

A simple \bd scattering amplitude with a sharp,
step-function edge is physically
implausible, even if the black disc picture is basically correct.
It seems more plausible to us to assume a fixed, soft edge on the
disc, where the  edge has {\em fixed,  \yy
independent} properties. This edge  will then  gradually become
relatively less important as the disc grows in size.

\subsection{Test for an Energy-Independent  Edge}
  We test  the picture of a finite, energy-independent edge in the
$pp$ scattering amplitude as
follows.

\begin{figure}[h]
\includegraphics[width=0.8\linewidth]{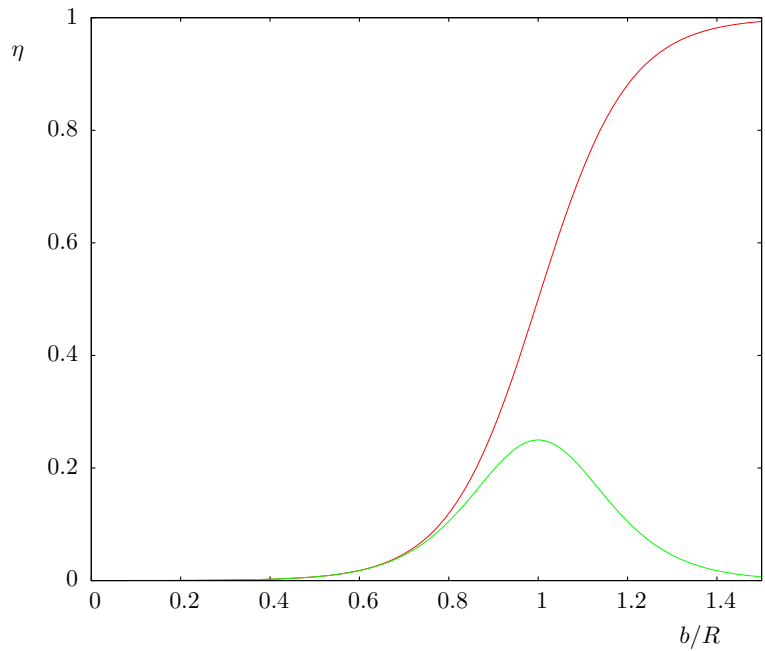}
\caption{Schematic shape of the  transparency $\eta$ as  a
function
of impact parameter
$b$ (upper curve, red). With a sharp edge, $\eta$ would simply jump
to $1$ at $b/R=1$. 
Also shown is the corresponding $\eta(1-\eta)$ (lower curve,
green).}
\label{etasketch}
\end{figure}

We assume initially that at high energies,  the elastic \sctg
amplitude is
purely imaginary. We
can then write the total and elastic cross sections in
two-dimensional  impact
parameter space $b$   \cite{block_rev} as
\beql{csss}
\sigma^{TOT}=8\pi\int_0^\infty
\hf (1-\eta)\,b\,db,~~~~~~~~~~~~~\sigma^{EL}=8\pi\int_0^\infty
(\hf(1-\eta))^2b\,db.
\eeql
  The quantity $\eta(b,s)$  is the ``transparency''  at impact
parameter $b$. In an eikonal picture it is given by the eikonal
function
$\chi(b,s)$ as $\eta(b)={\rm exp}(-\chi)$ \cite{block_rev}. It has
the general form indicated  in Fig.\,\ref{etasketch},
%%%%%%%%%%%%%%%%%%%%%%%%%%%%%%%%
%%%%% FIG. 1 %%%%%%%%
starting from approximately zero
at $b=0$ and rising rapidly to 1 in the vicinity of the black disc
radius $b=R$. If the
jump to 1 at $b=R$
were simply a step function, one would have the ideal black disc
with radius $R$
and  a sharp edge.
However with  a smooth rise as we have indicated in the figure,
there is an
edge, whose properties we wish to study.

A quantity which  exhibits the nature of the edge is
the following:
\beql{csssa}
\sigma^{TOT}-2\sigma^{EL}=4\pi\int_0^\infty
 \eta (1-\eta)\,b\,db \,.
\eeql
We note that that $\eta (1-\eta)$ vanishes both at $b=0$ and
$b=\infty$ while peaking near $b=R$ and therefore seems to be a
suitable quantity for
isolating the edge.

 If we now assume that the important contribution to the integral
in \eq{csssa} occurs over a relatively narrow range of $b$ around
$R$,
we can set the factor $b$ in the integrand to $R$ and write
\beql{csssb}  
\sigma^{TOT}-2\sigma^{EL}\approx 4\pi R\, {\cal I} \,,
\eeql
where ${\cal I}$ is the  integral  $\int_0^\infty
 \eta (1-\eta)\,db $. If the edge has energy-independent
properties, we
expect $\cal I$ to be constant.

Since the disc scattering dominates the cross section at high
energies,
 we approximate $R$ as $R=\sqrt{\sigma^{TOT}/2\pi}$. We thus expect
for  a constant edge  that

\beql{csssc} 
(\sigma^{TOT}-2\sigma^{EL})/\sqrt{(\pi/2)\sigma^{TOT}}\approx
4\, {\cal
I}\to {\rm constant} 
\eeql
at high energies. Since the maximum value of $\eta (1-\eta) $ is
$1/4$,
one may think of $\cal I$ as  ${\cal I}=\frac{1}{4} t$, where $t$
is an effective `thickness' of the edge. With this definition, the
ratio in \eq{csssc} is simply $t$.

\begin{figure}[h]
\includegraphics[width=\linewidth]{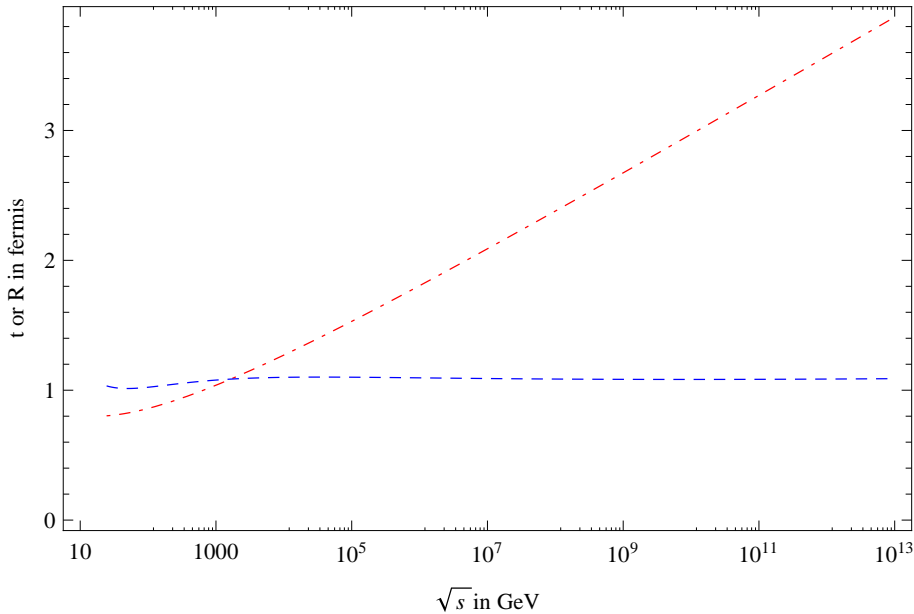}
\caption{ Plot of the ratio \eq{csssc}. The dashed (blue)
line represents $t$, the effective thickness  of the edge in
fermis, as explained in the text. For
comparison the dashed-dotted line (red) represents  the black-disc
radius $R$ inferred from the total \csssn, namely
$R=\sqrt{\sigma^{TOT}/2\pi}$, in fermis. }
\label{ratio}
\end{figure}

Figure\,\ref{ratio} shows an evaluation of the left-hand side  of
\eq{csssc}
using  the even combination of \csssn s
 $\hf(pp + \bar{p}p)$ from the preexisting fit
 from Ref.\,\cite{bh0}. This fit did not include data above $\sim
1.8$ TeV, but successfully
 and accurately predicted the cross sections measured at  LHC and
cosmic ray energies.  The dashed (blue) line
represents $t$.  For comparison the dashed-dotted line (red)
represents  the radius inferred from the total \csssn, namely
$R=\sqrt{\sigma^{TOT}/2\pi}$.

 The thickness $t$ is
approximately 1 fermi, and is quite constant over an enormous
energy region.

In defining $R$, one could also try  using
$R=\sqrt{\sigma^{EL}/\pi}$.
 This  does not make a large quantitative
difference, and leads to  the same asymptotic value for
$t$. But due to the slow logarithmic behavior of the
quantities the crossover is moved to  higher \yyn, over 100 TeV.

%%%%%%%%%%%%%%%%%%%%%%%%%%%
%%%%%%%% FIG. 2 %%%%%%%%%%%%%%%%
%%%%%%%%%%%%%%%%%%%%%%%%

\subsection{Real Part of the Amplitude}
 In writing the expressions in \eq{csss}, we have taken the elastic
scattering amplitude to be purely imaginary
and neglected its real part.  A  small real part is known to be 
present in the forward scattering amplitude $f(s,0)$ at high \yyn,
both from  direct measurements and from dispersion relations
\cite{bh1}. It reaches a peak of about 10\% of the imaginary 
part for energies $\sqrt{s}$ in the range 100-1000 GeV, and is
smaller at lower and higher energies.
Since our relation \eq{csssc} deals with \csssn s, where the real
part enters squared, we may expect a real part correction to be
small.

For example, in the eikonal representation of the scattering
amplitude \cite{block_rev}, where one
introduces an imaginary as well as real part to the eikonal
function, $\chi = \chi_R + i\chi_I$, one finds that
\beql{fRefIm}
{\rm Re}f(s,0) = -\int_0^\infty \eta \sin{\chi_I}\, b\, db, \qquad
{\rm Im}f(s,0) = \int_0^\infty (1-\eta\cos{\chi_I})\,b\,db,
\eeql
where $\eta=\exp{(-\chi_R)}$ is the transparency. The corresponding
edge integral  is
\beql{newedge}
\sigma^{TOT}-2\sigma^{EL}=4\pi\int_0^\infty
 \eta (\cos{\chi_I}-\eta)\,b\,db.
\eeql

 Since $\eta\approx 0$ in the black disc region, and $\chi$ is 
expected to vanish rapidly at large $b$, we conclude from the
observed 
smallness of the forward real-to-imaginary ratio
that $\chi_I$ is itself small in the edge region. The factor
$(\cos{\chi_I}-\eta)=(1-\eta)-\frac{1}{2}\eta\chi_I^2+\cdots$ in
\eq{newedge} therefore differs from the factor $(1-\eta)$ in
\eq{csssa} only in second order in $\chi_I$. We therefore expect
corrections of at most a few percent to the results discussed in
the preceding section. These expectations are confirmed by a
numerical calculation  in this eikonal model \cite{block_rev},
where the change in the integrand of the edge integral \eq{csssc},
with and without  $\chi_I$, is barely perceptible.

\section{Interpretation} \label{int}
The flatness of the curve for $t$ in Fig.\,\ref{ratio}
 is impressive. It strongly supports
the picture of a proton-proton scattering amplitude  consisting of
a growing black disc with
a  constant smooth edge with an energy-independent shape.

 It should be kept in mind that the  data used for the fit of the
$pp$ and $\bar{p}p$ cross sections in  Ref.\,\cite{bh0} 
extended only up to $\surd s\sim 1.8$ TeV. This fit was used
unchanged in making the  
curves in Ref.\ \cite{bh}; these include the
new LHC and cosmic ray data and show the accuracy of the
predictions.
 Since the fit  gives a very good representation of the new
experimental data up to $\sim 80$  TeV,
our curves  up to this \yy  could just as well have been made
directly from the experimental data. However, for energies above 80
TeV,  the extrapolation could depend possibly
 on the functional forms used in making the fit.

The  essential feature of the fit which leads to the constancy of
the ratio  in
\eq{csssc} is that the leading $\ln^2s$ terms in $\sigma^{TOT}$ and
$\sigma^{EL}$ appear with coefficients in the ratio 2:1 as noted in
\cite{bh1}, and so cancel in the \csssn s. The next-to-leading 
terms are proportional to $\sim \ln s$, but with different
coefficients, so that
 the difference $(\sigma^{TOT}-2\sigma^{EL})$ is  proportional to 
$\ln s$. 
 This logarithm is effectively cancelled by the leading $\ln s$
from the square root of
$\sigma^{TOT}$ in the denominator, leaving a constant difference up
to
terms of order $1/\ln{s}$.

 It is  conceivable that some other next-to-leading
parameterization---provided it gives a good lower \yy fit and is
consistent with the
Froissart bound---could give a different asymptotic behavior for
\eq{csssc}. But, it
should be noted both that  $ \ln s$ terms are naturally the leading
subdominant terms generated in eikonal models 
in which $\chi$ decreases exponentially at large $b$, and that such
 terms are also implicit when the
leading term is $\sim \ln^2 s$ since $s$ must appear with a scale
factor $s_0$, and this is interchangeable with 
a $\ln s$ term.

Thus, in the simplest picture where all \csssn s have the
same leading black disc  behavior, one expects asymptotically
\beql{j}
\sigma^{TOT}_j= \beta\,
\ln^2\sqrt {s/s^j_0}+...~~~~~~~~~~~~\sigma^{EL}_j=\hf  \beta\,
\ln^2\sqrt{s/s^{j'}_0}+...     \,,
\eeql
where $\beta$ is universal, according to \cite{bh} $\beta=1.1$ mb,
but the $s_0$ depend on the particular
reaction,
reflecting differences in the scale where universal behavior sets
in. When subtracting two \csssn s as in \eq{csssc}, this
necessarily leads to a 
 $\ln s$ term: $\sigma^{TOT}-2\sigma^{EL}=
2 \beta\, \ln\sqrt{s/s^j_0}\,\ln \sqrt{s^{j'}_0/s^j_0} +...$.
 This said, it would  of course be helpful
 if even higher \yy
data were somehow to become available in order to extend and
confirm the fits \cite{bh1}.

 The estimate of the thickness of the  edge, about $ 1$ fm, arrived
at
 from the constant on the right-hand side 
of \eq{csssc}, is quite reasonable. Since $\pi (1\, {\rm
fm})^2\approx 30 \,{\rm mb}$ is on the order of 
  the low \yy  $pp$ \csss\!\!, $\sim 40$ mb at $\surd s =30$ GeV,
one might say that at low \yy the
$pp$ scattering amplitude is `all edge',  that is, dominated by
$pp$ interactions at impact parameters $b\approx 1$ fm. 
 Since  the disc-type behavior only sets in around 1 TeV and grows
slowly, the relatively large thickness
of the \ed\, means that the transition to disc domination occurs at
high \yyn, at least around $\surd s\sim 10$ TeV.
This  gives a natural explanation  
of the slow approach to ``asymptopia''.

 The question of the nature or physical constitution of the edge
raises some
interesting points. Is it the same for reactions with different
\plsn? If
the apparently universal behavior of cross sections at high \yy
originates in the
gluon field, one would suppose that the disc and its edge are
asymptotically the same for
all \pl species. On the other hand, our observation that at low \yy
the scattering amplitudes appear to be all edge might suggest that
the differences in
cross sections at low \yy are preserved to high \yy through
differences in the edge.
Since, as we have explained, the edge gives a subdominant
contribution to the
total \csssn s, this would not affect the universality of the
\csssn s themselves.  
Unfortunately, data for other \pl species are not obtainable 
as directly  as for
protons, where one has the LHC and cosmic rays, but it would be of
great interest if such information became available.

\section{Acknowledgments}
  All of the authors would  like to thank the Aspen Center for
Physics for
its hospitality in making our working group possible, where
this work was supported in part by the
National Science Foundation under Grant No. PHYS-1066293.
TJW is supported in part by Department of Energy grant, DE-
SC0011981 and
Simons Foundation Grant 306329.
FH's research was supported in part by the U.S. National Science
Foundation under Grants No.~OPP-0236449 and PHY-0969061 and by the
University of Wisconsin Research
Committee with funds granted by the Wisconsin Alumni Research
Foundation. We thank E. Seiler for help with the references.

\end{document}